\documentstyle[floats,prb,aps,preprint]{revtex} \begin{document}

\input{psfig}

\title{Flux Noise and Fluctuation Conductivity in Unfrustrated
Josephson Junction Arrays} 
\author{Ing-Jye Hwang\cite{*}, Seungoh Ryu and D. Stroud}
\address{
Department of Physics,
The Ohio State University, Columbus, Ohio 43210}

\date{\today}

\maketitle

\begin{abstract}

We study the flux noise $S_{\Phi}(\omega)$ and finite frequency 
conductivity $\sigma_1(\omega)$ in two dimensional unfrustrated Josephson 
junction arrays (JJA's), by numerically solving the 
equations of the coupled overdamped resistively-shunted-junction model 
with Langevin noise.  We find that $S_{\Phi}(\omega)\propto \omega^{-3/2}$ 
at high frequencies $\omega$ and flattens at low $\omega$,
indicative of vortex diffusion, while $\sigma_1 \propto \omega^{-2}$ 
at sufficiently high $\omega$.  Both quantities show clear evidence of
critical slowing down and possibly scaling behavior near the 
Kosterlitz-Thouless-Berezinskii (KTB) transition.  The critical slowing
down of $S_{\Phi}$, but not its frequency dependence, is in agreement with recent
experiments on Josephson junction arrays.

\end{abstract} \newpage

\def\a{\alpha}
\def\ybco{$\rm YBa_2Cu_3O_{6.95}\:$}
\def\ncco{$\rm Nd_{1.85}Ce_{0.15}CuO_4\:$}
\def\bscco{$\rm Bi_2Sr_2CaCu_2O_{8+\delta}\:$}
\def\other{(put others)}

\newdimen\unicodeptsize \unicodeptsize=12pt

\begingroup
\catcode`@=11 

\gdef\Uni#1:#2:#3:#4:#5<#6>
 {\leavevmode \hbox to#1\unicodeptsize
    {
\hss}}

{\catcode`p=12\catcode`t=12\gdef\uni@ff#1pt{#1}}
\gdef\unic@deptsize{\expandafter\uni@ff\the\unicodeptsize\space}

\endgroup
\def\myname#1{\unicodeptsize=#1pt%
\Uni1.08:24:24:-1:20
<01c1c001818c7ffffe01818001818001ff8001818000000c7ffffe0000000800100ffff8
0c18380c18300c18300ffff00c18300c18300c18300ffff001838003c0780f001ef80006>%
\thinspace 
\Uni1.08:24:24:-1:20
<01010001c1c001818c7ffffe018180019180011d0000180004186007fff0061860061860
06186006186006186c7ffffe00340000320000630000c1c00180e006007e180030600000>%
\thinspace 
\Uni1.08:24:24:-1:20
<07301c062018062218063e1a0c66ff0c46180c8698196c98183c9a3018ff703018582018
98431818831818030219ffff180780180f40181b6018333818631f1983061803001e0300>%
\thinspace 
}

Josephson junction arrays (JJA's) and thin-film superconductors 
are excellent model systems for studying 
vortex dynamics.  At zero magnetic field, such systems are believed to undergo
a Kosterlitz-Thouless-Berezinskii\cite{kosterlitz73,berezinskii70}
(KTB) transition at a temperature $T_{KTB}$.
At temperatures below $T_{KTB}$ the vortices and anti-vortices 
are bound into pairs, whereas above $T_{KTB}$ these pairs start to
unbind into unpaired vortices.  Such a phase transition is expected to affect
a variety of transport properties of such
systems.\cite{ambegaokar}  Measurements of both the $IV$ characteristics
and the inverse kinetic inductance consistent with the occurrence
of a KTB transition have been reported in both thin superconducting 
films\cite{film1} and superconducting arrays.\cite{array1}

A particularly sensitive probe for such vortex dynamics is the study
of flux noise.   Conventional transport properties such as the $IV$ 
characteristics, while sensitive to vortex dynamics, are typically
nonequilibrium measurements, requiring the application of an external
current.  By contrast, magnetic flux noise is typically measured at 
equilibrium, by placing a
superconducting quantum interference device (SQUID) over a portion of
the array or film.   Such a measurement is sensitive to equilibrium
fluctuations in the local vortex number of vortices within that area.

A number of groups have studied flux noise in
superconductors.  Several measurements have been carried out in
in high temperature superconducting films, including
\bscco\cite{rogers92} and \ybco.\cite{ferrari91}
Recently, Shaw {\it et al.\/}\cite{shaw96} have done noise experiments 
on overdamped JJA's consisting of superconducting Nb islands 
in a Cu film, greatly extending some earlier 
measurements by Lerch {\it et al.\/}\cite{lerch}  These experiments yield a range of behavior for the spectral
function $S_{\Phi}(\omega)$ of the flux noise, 
that is, the frequency Fourier transform of the 
flux-flux correlation function.
For example, \ybco\cite{ferrari91} and JJA's\cite{shaw96}
are found to have $S_{\Phi}(\omega) \propto \omega^{-1}$ at ``high'' 
frequencies (``high,'' in this context, meaning greater than about 10-1000 Hz),
while in \bscco\cite{rogers92} $S_{\Phi}(\omega) \propto \omega^{-3/2}$ at 
similar frequencies.

There have also been several theoretical studies
of flux noise in such systems.  Houlrik {\it et al.\/}\cite{houlrik94}
discussed the behavior of flux noise from a Coulomb gas analogy, and
calculated $S_{\Phi}(\omega)$ from a time-dependent Ginzburg-Landau
(TDGL) model.  At high frequencies, they found
$S_{\Phi}(\omega) \propto \omega^{-2}$. 
Gr\/{o}nbech-Jensen {\it et al.\/}\cite{gronbech92} studied a JJA with a
static magnetic field of 1/2 flux quantum per plaquette, 
using the so-called resistively shunted junction (RSJ) model 
including self-capacitance.  Their primary interest, however, 
was to find the voltage noise at finite external currents, 
with disorder in the islands' positions, rather than the flux noise itself.
Recently, Wagenblast and Fazio
\cite{wagenblast97} have studied the flux noise and scaling behavior
using an $XY$-model with an assumed local damping for the phases.
The local damping term corresponds to ohmic
resistance shunts coupling each superconducting grain to the ground. 
The resulting flux noise was found to be white for low frequencies,
varied as $\omega^{-2}$ at high frequencies, and as $\omega^{-1}$ 
at intermediate frequencies.  Very recently, Tiesinga 
{\it et al.\/}\cite{tiesinga97} have used both the
coupled RSJ model and a TDGL model to study flux noise numerically over a
relatively limited frequency range.  They
concluded that their TDGL results were closer to the experiment 
of Shaw {\it et al.\/}\cite{shaw96} than were the RSJ predictions.

\def\vdot#1#2{{\bf #1\cdot #2}}
\def\vcross#1#2{{\bf #1\times #2}}

In this paper, we carry out extensive calculations of flux noise in an
array of coupled overdamped Josephson junctions, using Langevin noise to
simulate the effects of temperature.  Our model is similar to that of
Tiesinga {\it et al.\/}\cite{tiesinga97}, 
but we study the real part of the frequency-dependent fluctuation conductivity
$\sigma_1(\omega)$ in addition to the vortex noise, and we calculate both 
over a considerably wider frequency regime.  
Our results for the flux noise show a clear 
signature of vortex diffusion above $T_{KTB}$, i.~e., 
$S_{\Phi}(\omega) \propto \omega^{-3/2}$ above a cut-off frequency
$\omega_v(T)$ which approaches zero near $T_{KTB}$.  $\sigma_1(\omega)$ is
found also to have a characteristic frequency $\omega_{\sigma}(T)$ which
approaches zero on either side of $T_{KTB}$.

The details of the RSJ model can be found in the literature.\cite{chung89} 
The current through a junction between two
superconducting islands $i$ and $j$ is assumed to consist of three 
contributions in parallel:
a normal current $I_{R;ij}={V_{ij}/ R_{ij}}$ through 
a resistance $R_{ij}$; a Josephson 
current $I_{S;ij}=I_{c;ij}\sin(\theta_{ij})$; 
and a thermal noise current $I_{L;ij}$.
Here $I_{c;ij}$ is the critical current,  
$\theta_{ij}=\theta_{i}-\theta_{j}$ is the phase
difference across the junction, and  
$V_{ij}\equiv V_i-V_j$ is the voltage between islands $i$ and 
$j$.  The use of Kirchhoff's law for current conservation and of the
Josephson relation $V_{ij}=(\hbar/2e)(d\theta_{ij}/dt)$ leads to
a set of coupled first-order nonlinear differential equations.
We solve these coupled equations numerically by a standard algorithm
for square lattices of several sizes, as discussed, for example, 
by Chung {\it et al.\/}\cite{chung89}  We assume no external
current, and use periodic boundary conditions for an $N \times N$
square lattice of size $N_s=N\times N$ with no disorder, i.~e., 
$I_{c;ij} =I_c$ and
$R_{ij}=R$ for all $i,j$, and no external magnetic field.
The time iteration is accomplished using a
second-order Runge-Kutta procedure with
time intervals of order $dt=0.01 \tau_0$, 
where $\tau_0 = \hbar/(2eRI_c)$ is the characteristic time of the problem.
By means of the assumed Langevin dynamics, we can calculate not 
only time-dependent quantities, but also various equilibrium 
quantities\cite{parisi}, computed as time averages.  In the following, 
our results are presented in the 
``natural units'' of the problem.  Thus, the natural unit of time 
is $\tau_0,$
and of frequency, $\omega_0=1/\tau_0$.  Energy and temperature are 
given in units of $ \hbar I_c /(2e)$.

Many of the physical observables of interest are spectral functions,
that is, the Fourier transforms of various time correlation functions.  
For an observable $O(t)$, the spectral function is defined as
\begin{equation}
S_O(\omega)={1\over N_s}\times
\lim_{\Theta\to\infty}{1\over\Theta}
\left\vert\int_{-{\Theta/2}}^{\Theta/2} O(t)
e^{-i\omega t}\,dt\right\vert^2.
\end{equation}
With this definition, $S_O(\omega)$ has dimensions $[O^2 \cdot t]$, 
where $t$ denotes time.

\begin{figure}[tbp]
\centerline{\psfig{figure=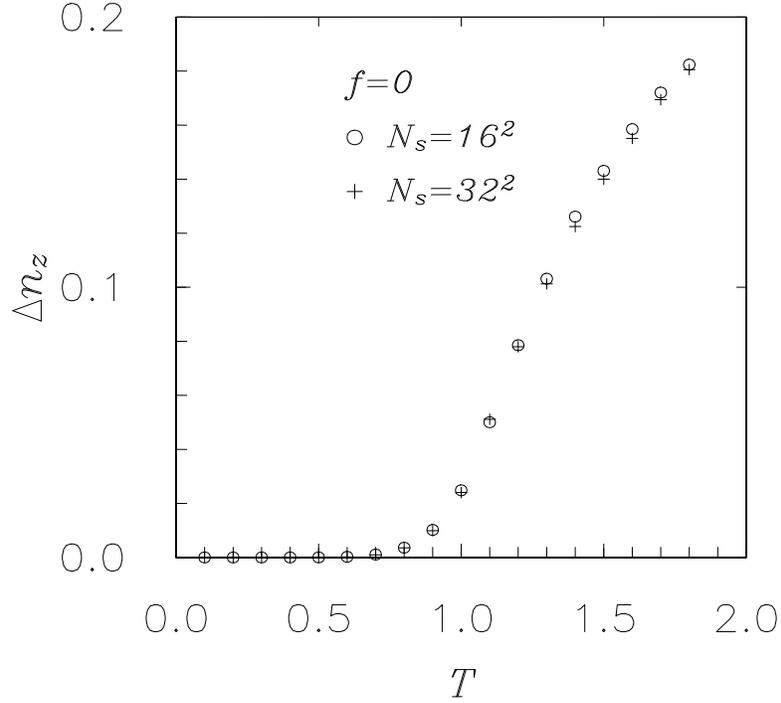,width=4in}}
\vskip 15pt
\caption
{
Time-averaged mean-square vortex density $\Delta n_z$ for the overdamped 
array at zero magnetic field, plotted as a function of temperature $T$ 
for two different array sizes.
\label{fg:szz_nz}
}
\end{figure}

Fig.~\ref{fg:szz_nz} shows the time-averaged mean-square vortex density 
$\Delta n_z=(1/N_s)\sum_{{\bf r}} n_z^2({\bf r})$ for zero magnetic field.
Here $n_{z}({\bf r}) = 0$ or  $\pm 1$ denotes the number of vortices in the 
${\bf r}^{th}$ plaquette, as conventionally defined.\cite{hwang96}
$\Delta n_z$ becomes nonzero near $T_{KTB}$, where the 
vortex-antivortex pairs begin to unbind.
$T_{KTB}$, as estimated from the vanishing of $\Delta n_z$, is close to the 
accepted value of $0.9$-$0.95\hbar I_c/(2ek_B)$ for an infinite 
lattice.\cite{teitel83}

\begin{figure}[tbp]
\centerline{\psfig{figure=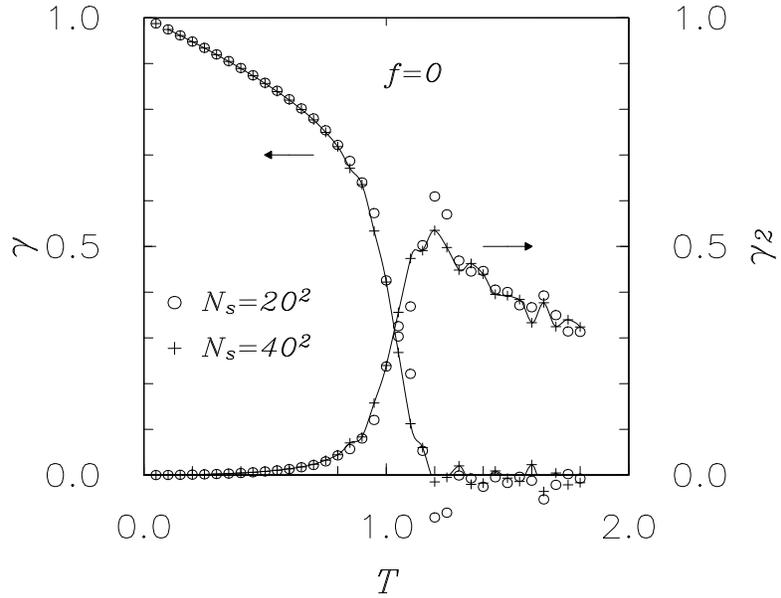,width=4in}}
\vskip 15pt
\caption
{
	Helicity modulus $\gamma$ and integrated fluctuation
conductivity $\gamma_2$ for an overdamped array at zero
magnetic field and two different array sizes, as calculated from
time-averaged solutions to RSJ equations with Langevin noise.
\label{fg:gxx}
}
\end{figure}

Another quantity of interest is the helicity modulus 
$\gamma$,\cite{fisher73} which measures stiffness 
against long-wavelength twists of $\theta$, and is proportional 
to the superfluid density.  Fig.~\ref{fg:gxx} shows $\gamma$ and the 
integral of the fluctuation conductivity, 
$\gamma_2=(1/\pi)\int_{-\infty}^\infty \sigma_1(\omega)d\omega$, where 
$\sigma_1=$ Re$\sigma$. 
Both are calculated using standard expressions,\cite{ebner83}
but from a time-average of the RSJ solutions.
$\gamma_2$ shows a characteristic peak near $T_{KTB}$.  The expected 
universal jump in
$\gamma$ at $T_{KTB}$ is somewhat broadened in our calculations, probably
by finite-size effects.

\begin{figure}[tbp]
\setbox0=\vbox{\hbox{(a)}\psfig{figure=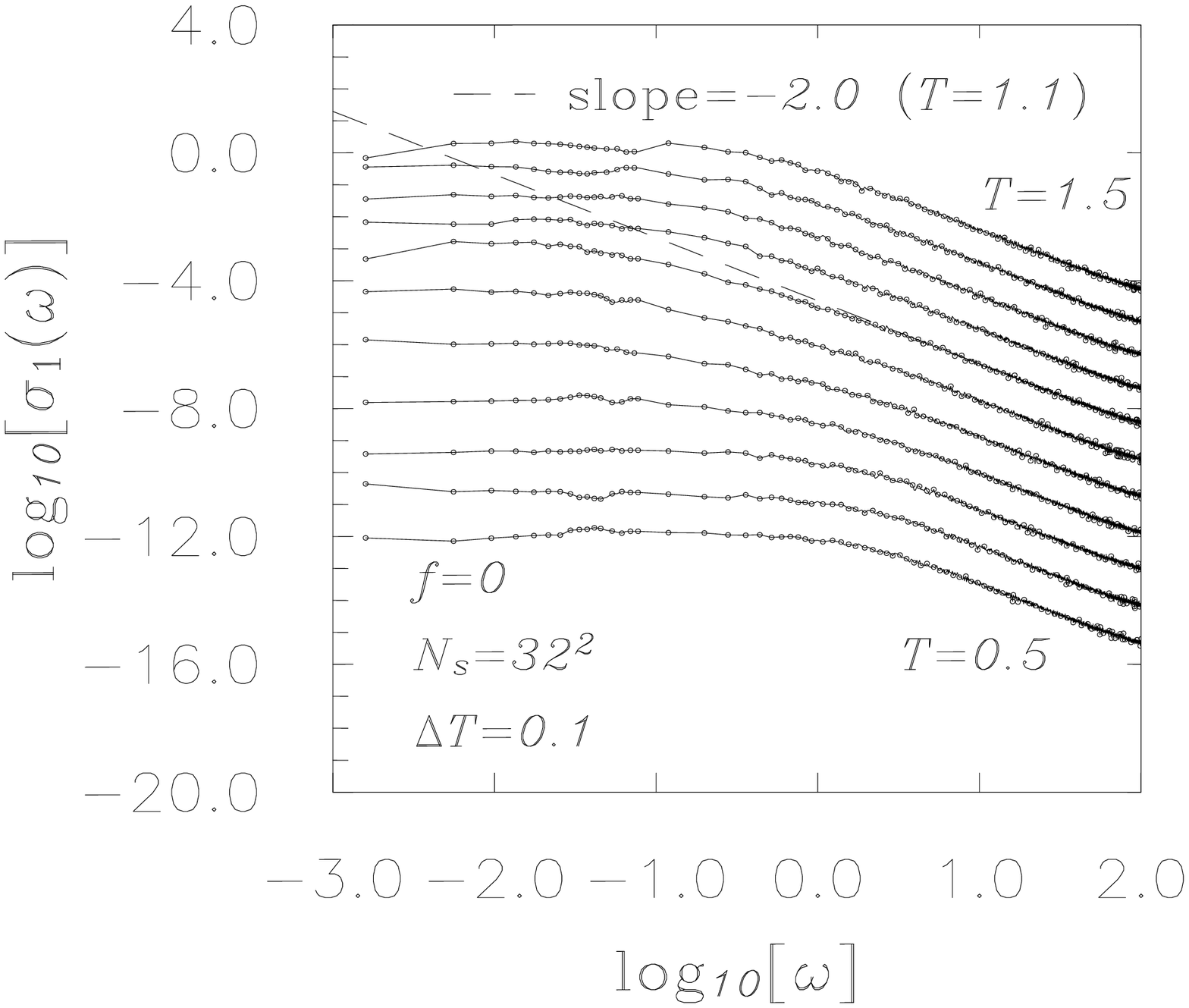,width=3in}}
\setbox1=\vbox{\hbox{(b)}\psfig{figure=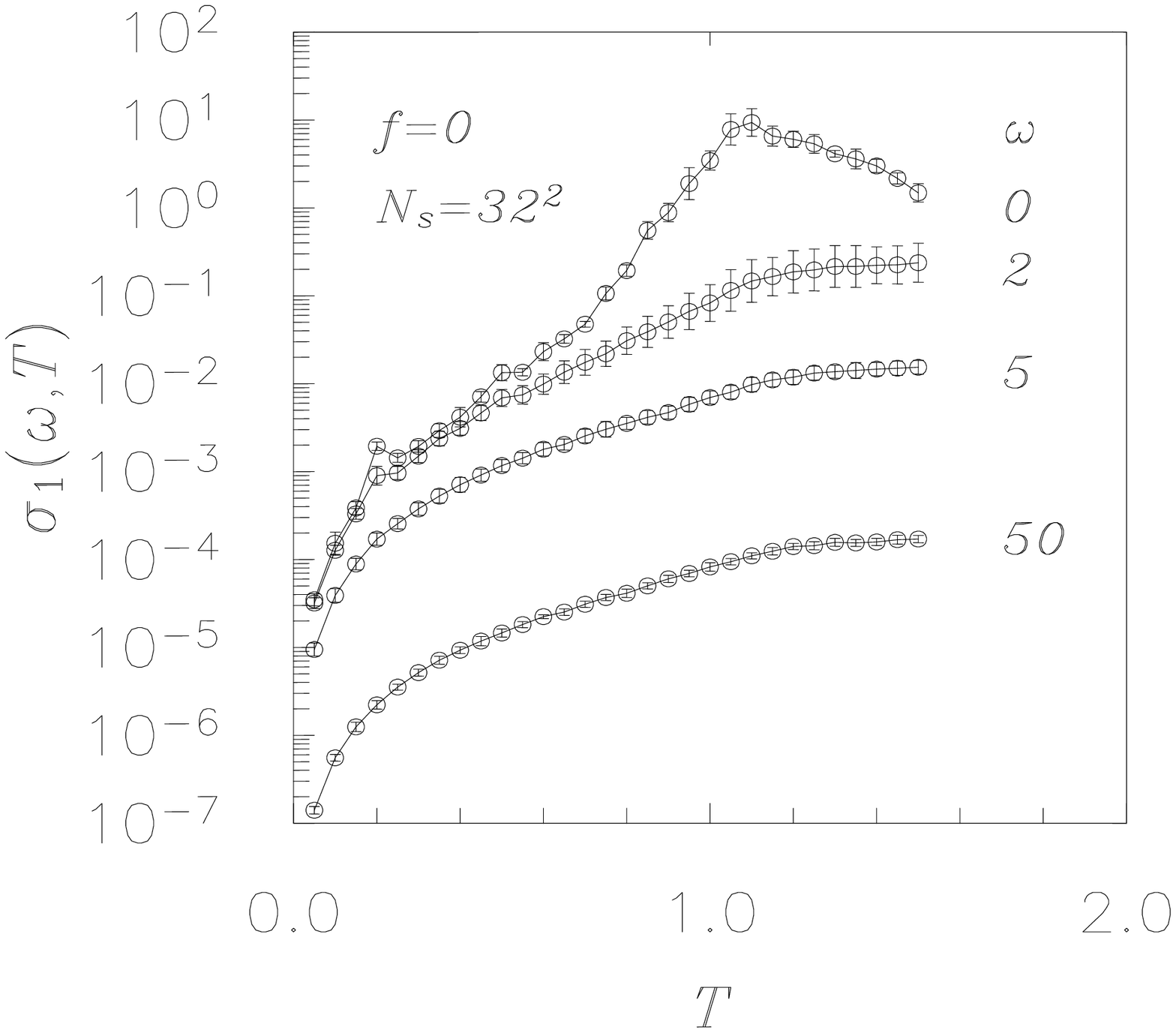,width=3in}}
\centerline{\box0\quad\box1}
\vskip 15pt
\caption
{
	Fluctuation conductivity $\sigma_1(\omega)$ (eq.\ \ref{eq:jxjx}), 
plotted (a) versus frequency $\omega$ for several temperatures $T$, and (b)
versus temperature at several $\omega$.  All quantities are plotted in natural
units.  The curves in (a) are vertically displaced; the dashed line has
slope $-2$.
\label{fg:jxf0}
}
\end{figure}

Fig.~\ref{fg:jxf0}(a) shows  $\sigma_1(\omega)$ itself, as calculated
directly from the fluctuation-dissipation theorem\cite{landau5}
for several different temperatures.
In the classical limit ($\hbar\omega \ll k_B T$) for this isotropic
system, this theorem gives
\begin{equation}
\sigma_1(\omega) = \frac{1}{N_sk_BT}\sum_{\bf r,r'} 
\int_0^{\infty} dt \cos(\omega t)
\langle J_x({\bf r}, t)J_x({\bf r}^{\prime}, 0)\rangle,
\label{eq:jxjx}
\end{equation}
where  
$
J_x = \sum_{\langle ij \rangle}I_c\sin(\theta_i-\theta_j)
$
is the $x$-component of the supercurrent. 
For a fixed temperature, $\sigma_1(\omega)$ flattens at low frequency and 
falls off at high frequencies approximately as $1/\omega^2$.
The slight upward
convexity at high frequency here (and for the noise calculation below)
is an artifact of the fast Fourier transform used to evaluate 
these quantities.  Fig.~\ref{fg:jxf0}(b) shows $\sigma_1(\omega,T)$ 
at several fixed $\omega$'s.  At the lowest frequency (nominally
$\omega = 0$, but actually an average over several $\omega < 0.08$),
there is a strong peak at $T \approx T_{KTB}$ (somewhat masked by the
log-log plot).  For the other frequencies (all greater than $\omega_0$), no
peak is discernable near $T_{KTB}$, indicating that the influence of the
transition is suppressed at such high frequencies.  
By comparing these results with those of Fig.~\ref{fg:gxx}, 
we see that the peak in $\gamma_2$ is dominated by the
{\em low frequency} regime of $\sigma_1(\omega)$, i.~e., $\omega < \omega_0$.

\begin{figure}[tbp]
\setbox0=\vbox{\hbox{(a)}\psfig{figure=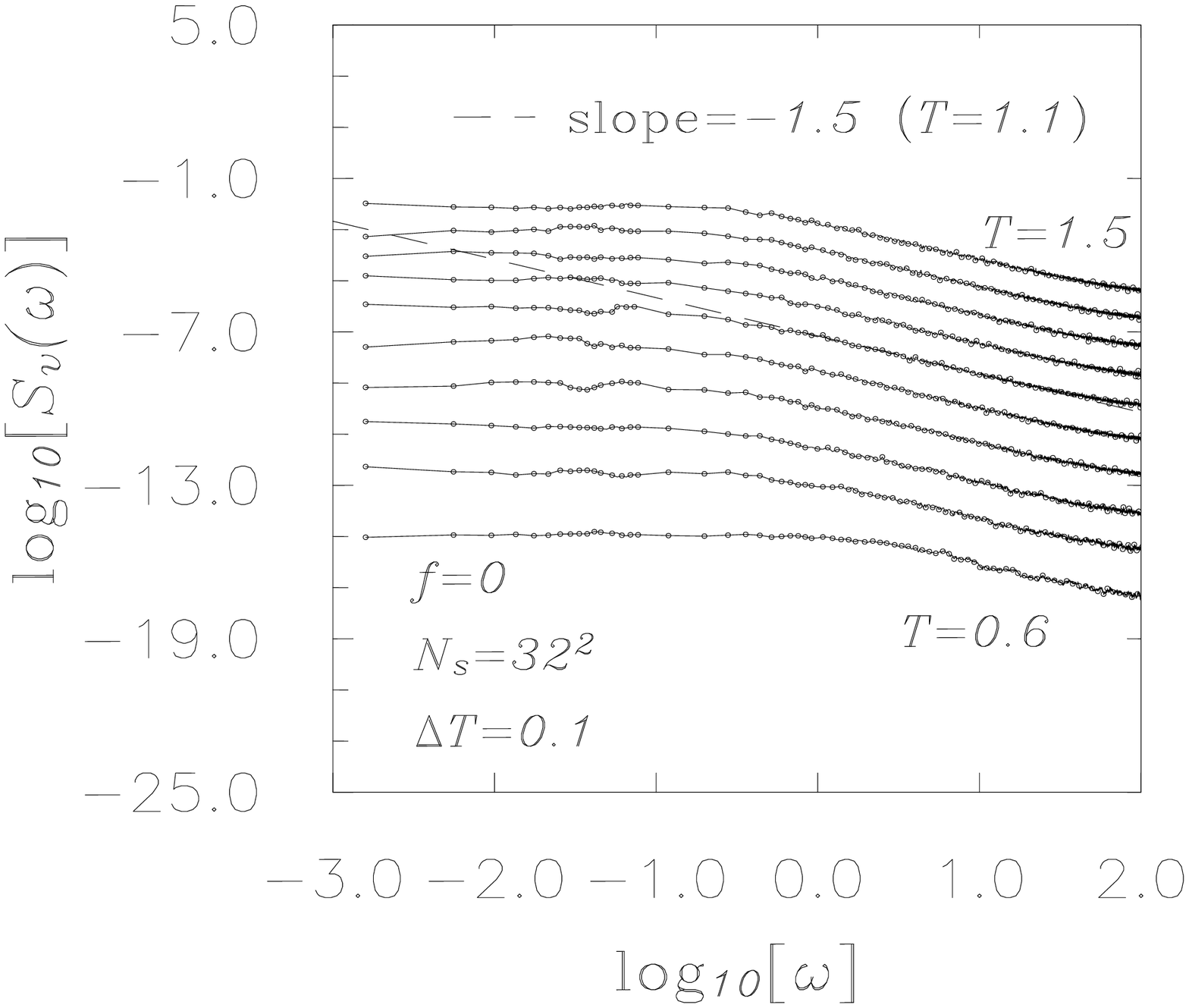,width=3in}}
\setbox1=\vbox{\hbox{(b)}\psfig{figure=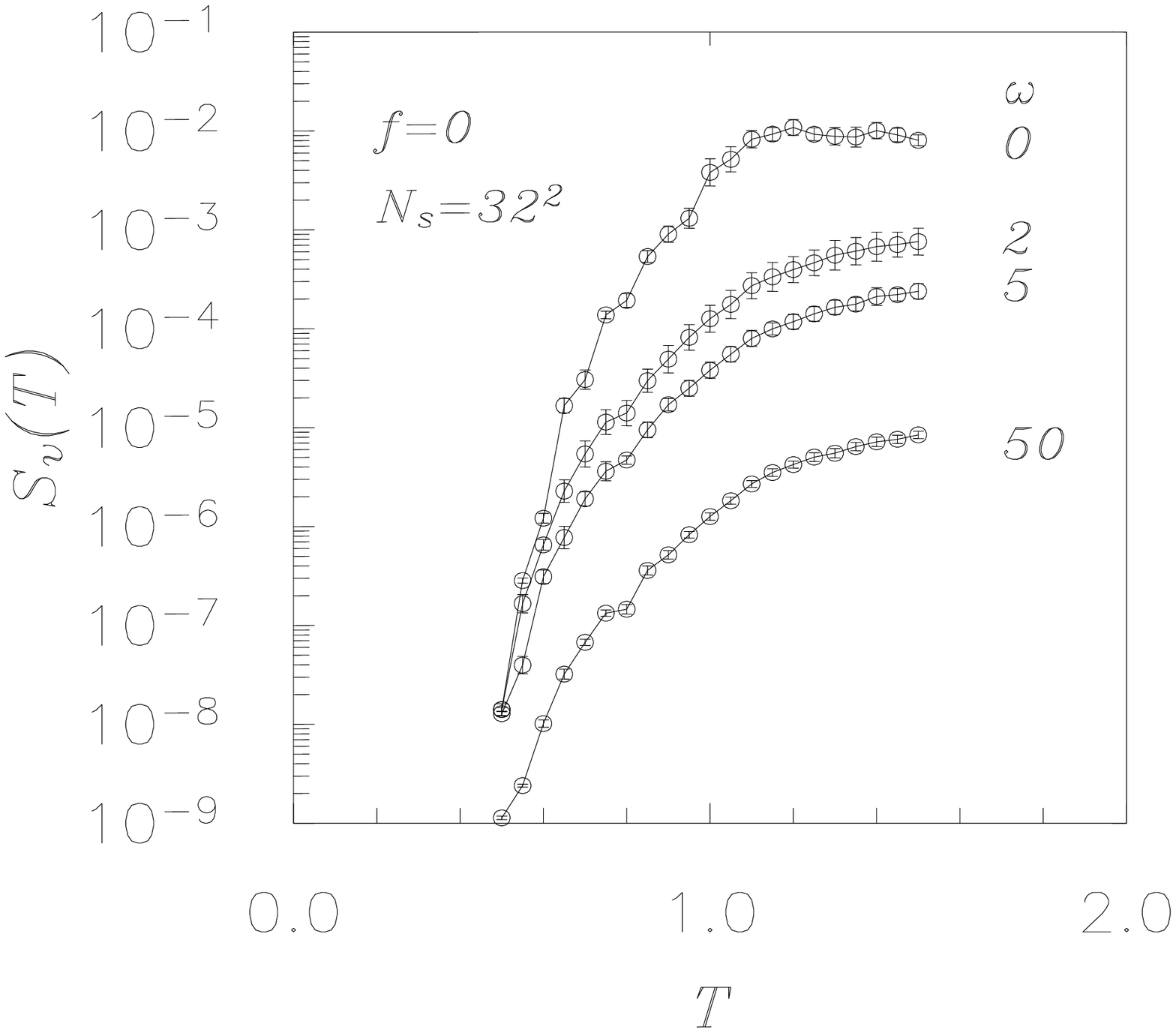,width=3in}}
\centerline{\box0\quad\box1}
\vskip 15pt
\caption
{
	Same as Fig.~\ref{fg:jxf0}, but for $S_v(\omega)$.  Dashed line has slope $-1.5$.  
\label{fg:nzf0}
}
\end{figure}

Next, we turn to the flux noise $S_{\Phi}(\omega)$.
Rather than calculate this quantity, we instead calculate the somewhat simpler
{\em vortex number noise}, $S_{v}(\omega)$, which should behave
similarly in most cases.\cite{shaw96}  $S_v(\omega)$ is the noise associated
with
$
N_v = \sum_{\bf r}n_{z}({\bf r}),
$
the vortex number enclosed in the area ${\cal A}$ 
spanned by the SQUID used to detect the flux noise.
With periodic boundary conditions, $N_v$ would be identically zero
if ${\cal A}$ were the entire array area.  For a smaller ${\cal A}$,
$N_v$ fluctuates in time, giving rise to vortex noise.   
Our results are shown in Fig.~\ref{fg:nzf0} for a 
square area ${\cal A} \equiv \ell^2$ equal to 1/4 of the array area.
Fig.~\ref{fg:nzf0}(a) shows our results for different temperatures as a
function of frequency; (b), for different frequencies as a function of $T$.
At low frequencies, $S_v$ becomes roughly frequency-independent, 
but at high frequencies $S_v(\omega) \sim  \omega^{-3/2}$, a dependence
which is known to characterize {\em diffusive behavior}.\cite{voss76}  
The spectral function thus clearly shows the characteristic signature of 
{\em vortex diffusion} at temperatures above $T_{KTB}$.

\begin{figure}[tbp]
\setbox0=\vbox{\hbox{(a)}\psfig{figure=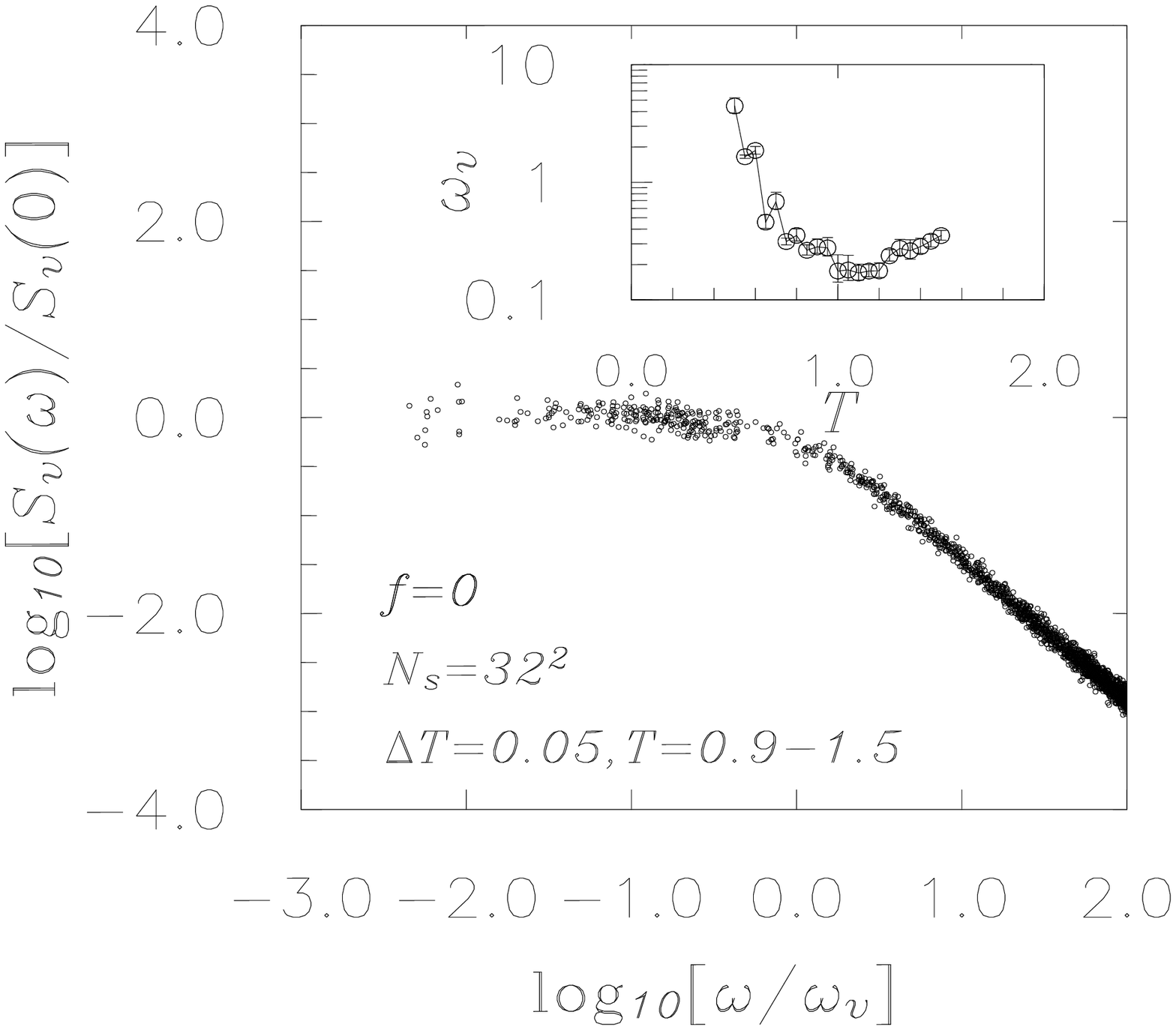,width=3in}}
\setbox1=\vbox{\hbox{(b)}\psfig{figure=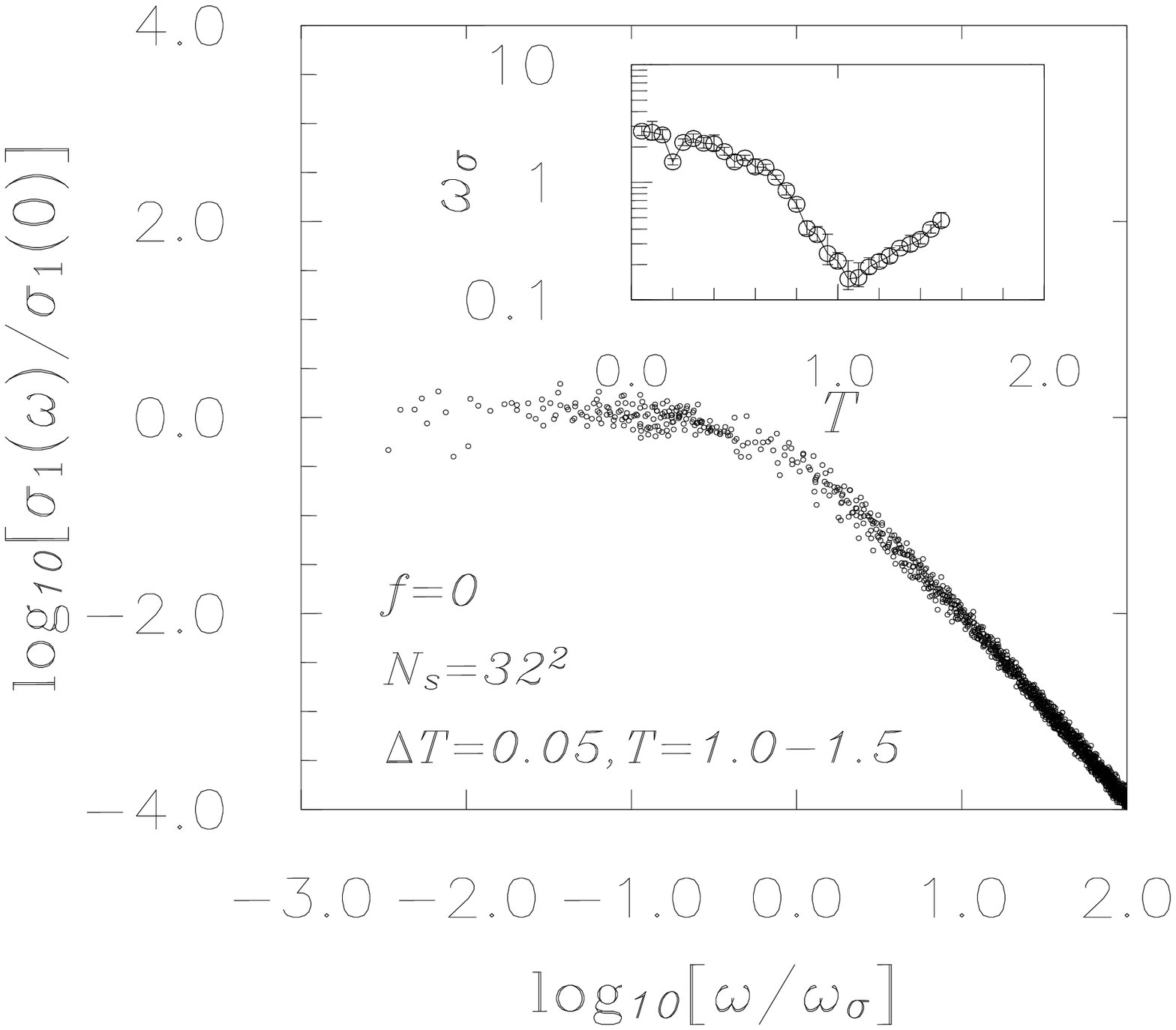,width=3in}}
\centerline{\box0\quad\box1}
\vskip 15pt
\caption
{
(a)
$S_v(\omega,T)$/$S_v(0, T)$ plotted against  $\omega/\omega_v$, where
$\omega_v(T)$ is a crossover frequency extracted from the data at each
temperature and plotted in the inset.
(b). $\sigma_1(\omega, T)/\sigma_1(0, T)$ plotted against
$\omega/\omega_{\sigma}$, where $\omega_{\sigma}(T)$ is plotted in the inset.  
\label{fg:scale}
}
\end{figure}

Fig.~\ref{fg:nzf0}(a) also suggests the same critical slowing down seen
in the conductivity plots.  That is, the $S_v(\omega)$ plots flatten
out at a temperature-dependent crossover frequency $\omega_{v}(T)$ 
[inset of Fig.~\ref{fg:scale}(a)]
which vanishes as $T \rightarrow T_{KTB}$ from either side.  
Such behavior suggests
that the effective vortex diffusion coefficient $D \rightarrow 0$ 
as $T \rightarrow T_{KTB}$.  This follows from the approximate
relation\cite{voss76} $D \approx \ell^2\omega_{v}$ which connects the $D$ to
the vortex noise spectrum in a fixed area $\ell^2$.

We now discuss more quantitatively the apparent critical slowing down seen in both our
flux noise and conductivity calculations.  In the former case, 
Shaw {\it et al.\/}\cite{shaw96} have suggested a scaling form for 
$S_{\Phi}(\omega)$ which may be written 
\begin{equation}
\omega S_{\Phi}(\omega) = F\left(\frac{\omega}{\omega_{\xi}}\right),
\end{equation}
where $\xi$ is the divergent correlation length characterizing the
phase transition, and $F$ is an appropriate scaling function (we ignore
any dependence on other possible scaling variables such as $\ell/\xi$).  
The frequency $\omega_{\xi}$ is expected to vary as $\xi^{-z}$, where
$z$ ($\approx 2$ experimentally\cite{shaw96}) is a dynamical critical exponent.
A simple scaling function which appears consistent with our calculations 
for $S_v(\omega)$ is
\begin{equation}
F(x) = \frac{x}{(1+x)^{3/2}}.
\end{equation}
As a partial test of this form, we plot in Fig.~\ref{fg:scale}(a) the ratio 
$S_v(\omega)/S_v(0)$ against $\omega/\omega_{v}$, where $\omega_v$ is
a crossover frequency used as a fitting parameter to the numerical data.
Clearly, all the plots fall atop one another, consistent with the
scaling hypothesis, and are in good agreement with the form (4), 
which gives the diffusive form $S_v(\omega) \propto \omega^{-3/2}$ at high
frequencies.  The crossover frequency $\omega_v$ [inset of 
Fig.~\ref{fg:scale}(a)]
appears to go to zero near
$T_{KTB}$ as required by the scaling form, but our data are not adequate to
test the expectation\cite{shaw96} 
$\omega_v = \omega_{\xi}\propto \xi^{-2}$ with 
$\xi \propto \exp\left[b\sqrt{T_{BKT}/(T-T_{BKT})}\right]$. 
In short, our numerical results for $S_{v}$ at $T > T_{KTB}$ are consistent
with diffusive vortex motion in the RSJ model, and 
a diffusion coefficient which vanishes continuously as $T \rightarrow T_{KTB}$.

The fluctuation conductivity $\sigma_1(\omega)$ exhibits similar critical
behavior, since it has the form of a Drude peak whose width goes
continuously to zero as $T \rightarrow T_{KTB}$.   We may anticipate
a scaling form $\sigma_1(\omega) = \sigma_0/(\omega^2+\omega_{\sigma}^2)$, where 
$\sigma_0$ has critical behavior at $T_{KTB}$; this form correctly gives the calculated
$1/\omega^2$ high-frequency behavior.  As a partial
test of this hypothesis, we have plotted in Fig.~\ref{fg:scale}(b), the ratio 
$\sigma_1(\omega,T)/\sigma_1(0,T)$ versus $\omega/\omega_\sigma$ for several
temperatures, again using $\omega_\sigma$ as a fitting parameter for each
temperature [inset of Fig.~\ref{fg:scale}(b)].  
The calculated $\sigma_1(\omega,T)$ all fall on the same universal
curve both above and below $T_{KTB}$ (for clarity,
we have plotted only the temperatures above $T_{KTB}$).  Moreover, the crossover
frequencies $\omega_\sigma$, like 
$\omega_v$, vanish smoothly as $T \rightarrow T_{KTB}$ from
either side, though again our numerical data is inadequate to confirm the
functional form $\omega_\sigma \propto \xi^{-z}$ with $z \approx 2$.

Our results differ from the $1/\omega$ behavior
seen experimentally for S$_{\Phi}(\omega)$
in overdamped Josephson arrays\cite{shaw96} over several
decades of frequency, although we do see indications
of similar scaling behavior.   Measurements on \ybco\cite{ferrari91} 
show a similar $1/\omega$ frequency regime, but as noted earlier, studies of 
very thin films of \bscco\cite{rogers92} also 
show a $1/\omega^{3/2}$ behavior in the same (100 Hz-10KHz) frequency regime, 
which is also interpreted as evidence for vortex diffusion.  Our results show
clear evidence of such diffusive behavior in a numerical model for the
dynamics of a Josephson array; the diffusing objects in
our calculations, as presumably in the \bscco films, 
are the thermally excited vortices and antivortices.  
The frequencies where we find diffusive behavior appear, however, to lie
well above the diffusive regime in \bscco.\cite{rogers92}  

It remains unclear why our calculations do not show a clear $1/\omega$ regime, 
as seen in some experiments and reported in calculations using a local damping 
model.\cite{tiesinga97}  One possibility is
simply that we have not probed the noise spectrum to sufficiently low 
frequencies, or, possibly, in a sufficiently large area.  
In the experiments, the $1/\omega$ regime
occurs in the KHz regime, and certainly well below MHz.  By
contrast, our calculations do not probe frequencies much below
about 0.01$\omega_0$ = 0.01 $2eRI_c/\hbar$.  For typical array parameters,
$\omega_0$ may be in the range of MHz or higher, suggesting that it could
be quite difficult to approach the $1/\omega$ regime in such numerical 
calculations.  

In conclusion, we have calculated both the vortex number noise
$S_{v}(\omega)$ and the frequency-dependent conductivity $\sigma_1(\omega)$
in an overdamped Josephson-junction array near the Kosterlitz-Thouless
transition.  The former exhibits a $\omega^{-3/2}$ behavior, characteristic
of vortex diffusion, at high frequencies, and is flat at low frequencies.
The latter has a Drude peak and a $1/\omega^2$ frequency dependence at
high frequencies.  Both quantities show clear evidence of critical slowing
down (i.~e., a vortex diffusion coefficient which goes to zero near $T_{KTB}$)
and possible scaling behavior near the Kosterlitz-Thouless-Berezinskii 
transition, in agreement with experiment; but clear evidence of a 
1/$\omega$ regime for the vortex number noise is lacking.

This work was supported by 
NSF Grant DMR94-02131 and DOE Grant DE-FG02-90 ER45427 through the
Midwest Superconductivity Consortium. 
Calculation were carried out using the SP2 at the Ohio Supercomputer
Center.

\end{document}